\documentclass[twocolumn,floatfix,secnumarabic,amssymb,amsmath,aip,prb]{revtex4}

\usepackage{graphicx}
\usepackage{bm}

\begin{document}

\title{Selection Rules for One- and Two-Photon Absorption
by Excitons in Carbon Nanotubes}

\author{Eduardo B. Barros$^{1,9}$\email{ebarros@mgm.mit.edu}}
\author{Rodrigo. B. Capaz$^{2,3}$}
\author{Ado Jorio$^4$}
\author{Georgii G. Samsonidze$^9$}
\author{Antonio G. {Souza Filho}$^{1}$}
\author{Sohrab Ismail-Beigi$^5$}
\author{Catalin D. Spataru$^{6,7}$}
\author{Steven G. Louie$^{6,7}$}
\author{Gene Dresselhaus$^8$}
\author{Mildred S. Dresselhaus$^9$}

\affiliation{$^1$Departamento de F\'isica, Universidade Federal do
Cear\'a, Fortaleza, Cear\'a, CEP 60455-760 Brazil.\\
$^2$Instituto de F\'isica, Universidade Federal do Rio de Janeiro,
Caixa Postal 68528, Rio de Janeiro, RJ 21941-972, Brazil\\
$^3$ Divis\~ao de Metrologia de Materiais, Instituto Nacional de
Metrologia, Normaliza\c c\~ao e Qualidade Industrial - Inmetro, R.
Nossa Senhora das Gra\c cas 50, Xer\'em, Duque de Caxias, RJ
25245-020, Brazil\\
$^4$Departamento de F\'isica, Universidade Federal de Minas Gerais,
Belo Horizonte, MG,
30123-970, Brazil.\\
$^5$Department of Applied Physics, Yale University, New Haven, Connecticut 06520, USA\\
$^6$Department of Physics, University of California at Berkeley, Berkeley, CA 94720, USA\\
$^7$Materials Science Division, Lawrence Berkeley National
Laboratory, Berkeley, CA 94720, USA\\
$^8$Francis Bitter Magnet Lab, Massachusetts Institute of Technology, Cambridge, MA 02139-4307, USA\\
$^9$Departament of Electrical Engineering and Computer
Science,Massachusetts Institute of Technology, Cambridge, MA
02139-4307, USA}

\date{\today}

\begin{abstract}
Recent optical absorption/emission experiments showed that the lower
energy optical transitions in carbon nanotubes are excitonic in
nature, as predicted by theory. These experiments were based on the
symmetry aspects of free electron-hole states and bound excitonic
states. The present work shows, however, that group theory does not
predict the selection rules needed to explain the two photon
experiments. We obtain the symmetries and selection rules for the
optical transitions of excitons in single-wall carbon nanotubes
within the approach of the group of the wavevector, thus providing
important information for the interpretation of theoretical and
experimental optical spectra of these materials.
\end{abstract}

\date{\today}

\maketitle

The use of symmetry is crucial for the description of the optical
spectra of atoms, molecules and solids. In the case of single-wall
carbon nanotubes (SWNTs), it has been predicted that excitonic
effects are key to understand their optical
transitions.\cite{ando97,chang04,spataru041,perebeinos04,zhao04}
Recent works have used the symmetry aspects of different excitonic
states in carbon nanotubes to prove the excitonic nature of their
optical spectra.\cite{wang05,maultzsch05} However, an analytical
study of the symmetry of the excitonic states cannot be found in the
literature, and a detailed analysis of the selection rules for one
and two photon absorption has not yet been reported. Therefore, an
analysis of exciton symmetries in SWNTs is needed to understand in
greater detail many aspects of their optical properties. In this
work, we use group theory to obtain the symmetries of the excitonic
states in SWNTs, as well as the selection rules for optical
absorption and emission, for one- and two-photon excitation process.
We describe in detail the number and symmetries of exciton states
for chiral $(n,m)$, zigzag $(n,0)$ and armchair $(n,n)$ SWNTs. Our
group theory analysis show that the results of the 2-photon
absorption experiments cannot be explained by symmetry-related
selection rules. The results reported here should form a basis for
helping the interpretation of theoretical and experimental optical
spectra of SWNTs.

The symmetry of excitons is developed here within the formalism of
the group of the wavevector, which has been covered partially in the
literature \cite{alon03} and will be more fully developed in a
future publication.\cite{barrosreview} Briefly, the factor groups
for the wavevector $k$ at the center ($k=0$) and edge of the
Brillouin zone ($k=\pi/T)$ are isomorphic to the $D_N$ ($D_{2nh}$)
point group for chiral (achiral) nanotubes, while the factor group
for a general wavevector $k$ is isomorphic to the group $C_N$
($C_{2nv}$). Here $N$ ($2n$) denotes the number of hexagons in the
unit cell for chiral (achiral) nanotubes and $T$ is the length of
the real space unit cell. The irreducible representations of the
factor groups of nanotubes are labeled by the quasi-angular momentum
quantum number $\tilde{\mu}$ which varies between $1-N/2$ and
$N/2$.This quantum number $\tilde{\mu}$ is related to the projection
of the compound symmetry operation ($\{R|\tau\}$) in the
circumferential direction of the nanotube, and can be associated
with the concept of cutting lines \cite{t1008}. Another quantum
number, of course, is the wavevector $k$, related to translation
symmetry. There are also parity quantum numbers related to a $C_2$
rotation (a $\pi$ rotation perpendicular to the tube axis, bringing
$z$ to $-z$), reflections, and inversion
operations.\cite{barrosreview,damn00}

A different but equivalent formalism is based on line
groups\cite{damn99,damn00}. The connection between the two
formalisms can be obtained through Table~\ref{tsb} that, despite its
technical aspect, is presented here for a clear definition of the
symmetry-related quantum numbers in both group theory formalisms
used in the literature.

\begin{table}
\begin{footnotesize}
\caption{Irreducible representations ($\mathcal{D}$) relevant to the
exciton problem for chiral and achiral nanotubes. GWV and LG stand
for ``group of the wavevector'' and ``line group'' notations,
respectively. The dimension ($d$) of each representation is shown on
the right for both GWV and LG formalisms. The last column describes
the wavevector ($k$), quasi-angular momentum ($\tilde{\mu}$) and
parity quantum numbers ($\Pi$). For chiral tubes, the relevant
parity is related to the $C_2$ operation ($\Pi^{C_2}$), whereas for
achiral tubes the parity $\Pi$ is also related to $\sigma_h$,
$\sigma_v$ reflections and inversion $i$. The GWV notation chooses
the parity under $i$ as a quantum number and the LG notation chooses
the parity under $\sigma_h$ as a quantum number, thus making the
translation between the two notations somewhat cumbersome. A zero
parity quantum number means that the representation does not have a
well defined parity.\label{tsb}}
\begin{tabular}{cccccc}\hline\hline
       & GWV & & LG & & \\\hline
Chiral & $\mathcal{D}$ & $d$ & $\mathcal{D}$ & $d$ & $(k,\tilde{\mu},\Pi^{C_2})$\\
       & $A_1(0)$ & 1 & $_0A_0^+$ & 1 & $(0,0,+1)$ \\
       & $A_2(0)$ & 1 & $_0A_0^-$ & 1 & $(0,0,-1)$ \\
       & ($\mathbb{E}_{\tilde{\mu}}(k) + \mathbb{E}_{-\tilde{\mu}}(-k) )$ & 1 & $_kE_{\tilde{\mu}}$ & 2 & $(\pm k,\pm \tilde{\mu},0)$ \\
\hline
Achiral & $\mathcal{D}$ & $d$ & $\mathcal{D}$ & $d$ & $(k,\tilde{\mu},\Pi^{\sigma_v}, \Pi^{\sigma_h}, \Pi^i, \Pi^{C_2})$\\
       & $A_{1u}(0)$ & 1 & $_0B_0^-$ & 1 & $(0,0,-1,-1,-1,+1)$ \\
       & $A_{2u}(0)$ & 1 & $_0A_0^-$ & 1 & $(0,0,+1,-1,-1,-1)$ \\
       & $A_{1g}(0)$ & 1 & $_0A_0^+$ & 1 & $(0,0,+1,+1,+1,+1)$ \\
       & $A_{2g}(0)$ & 1 & $_0B_0^+$ & 1 & $(0,0,-1,+1,+1,-1)$ \\
       & $E_{|\tilde{\mu}|u}(0)$ & 2 & $_0E_{|\tilde{\mu}|}^{\Pi^{\sigma_h}}$ & 2 & $(0,\tilde{\mu},0,(-1)^{\tilde{\mu}+1},-1,0)$ \\
       & $E_{|\tilde{\mu}|g}(0)$ & 2 & $_0E_{|\tilde{\mu}|}^{\Pi^{\sigma_h}}$ & 2 & $(0,\tilde{\mu},0,(-1)^{\tilde{\mu}},+1,0)$ \\
       & $(B'(k)+B'(-k))$ & 1 & $_kE_n^A$ & 2 & $(\pm k,n,+1,0,0,0)$ \\
       & $(B''(k)+B''(-k))$ & 1 & $_kE_n^B$ & 2 & $(\pm k,n,-1,0,0,0)$ \\
       & ($E_{|\tilde{\mu}|}(k)+ E_{|\tilde{\mu}|}(-k)$) & 2 & $_kG_{\tilde{\mu}}$ & 4 & $(\pm k,\tilde{\mu},0,0,0,0)$ \\
\hline\hline
\end{tabular}
\end{footnotesize}
\end{table}

\begin{figure*}
\includegraphics[width=5.2cm]{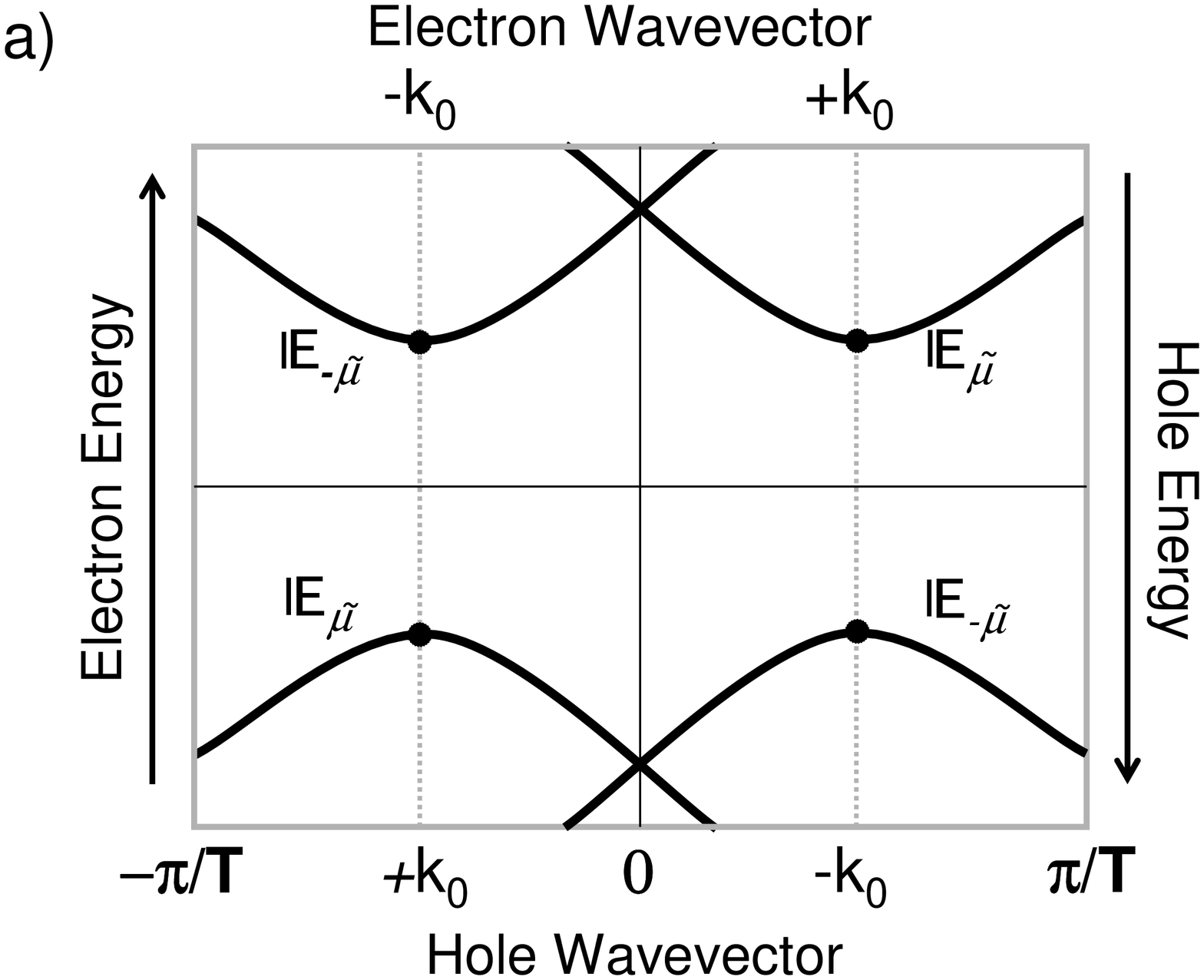}
\includegraphics[width=5.2cm]{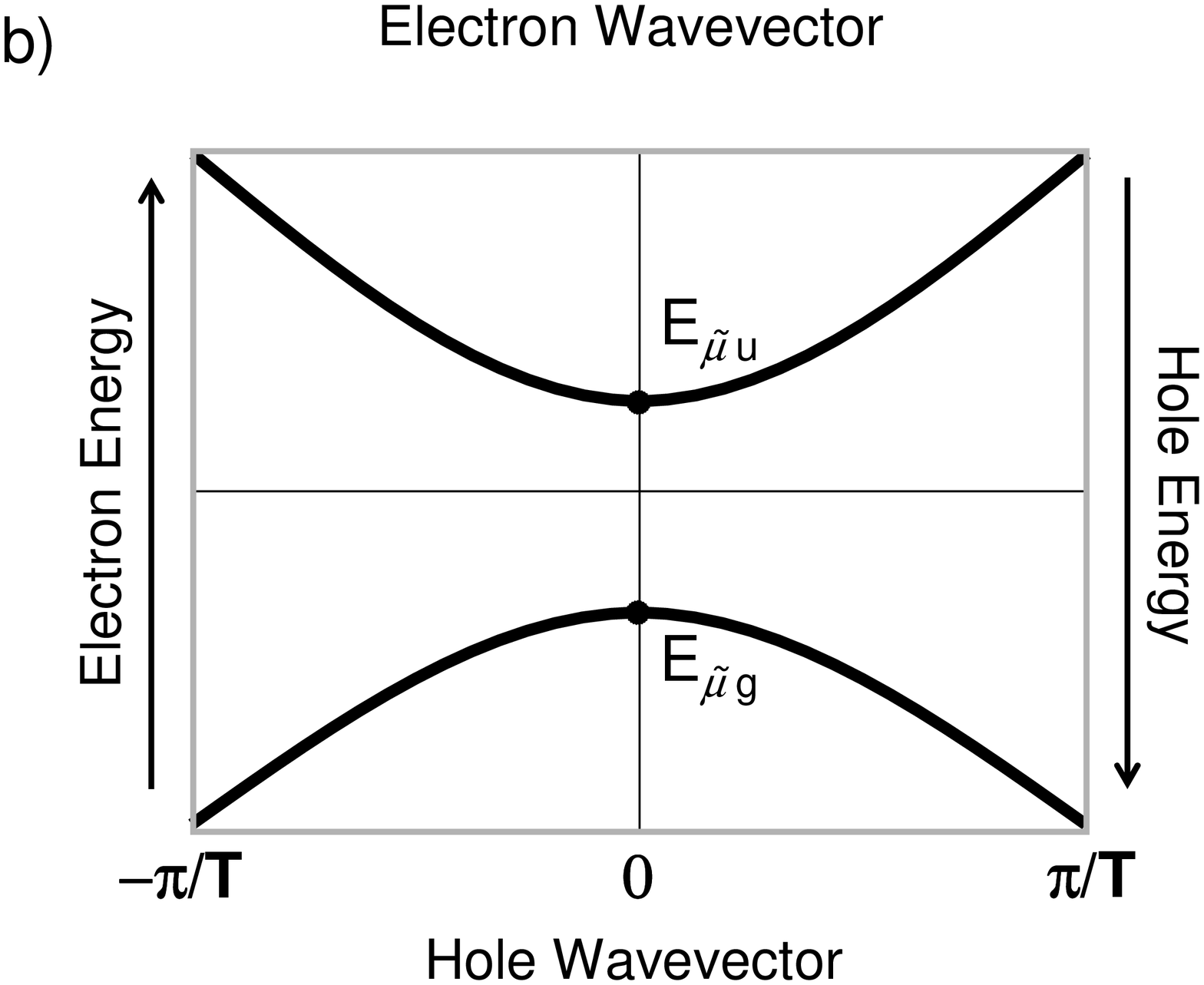}
\includegraphics[width=5.2cm]{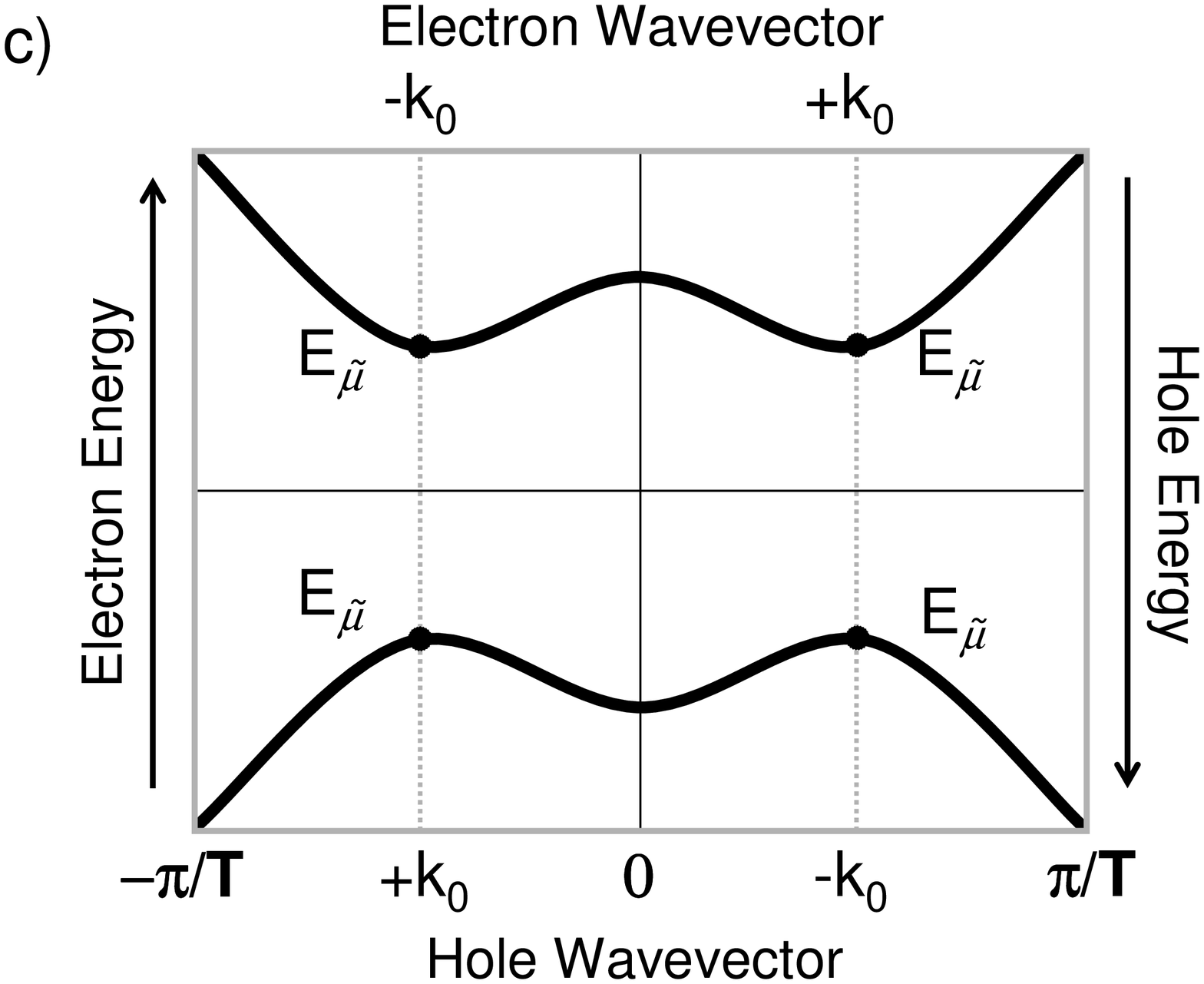}
\includegraphics[width=5.2cm]{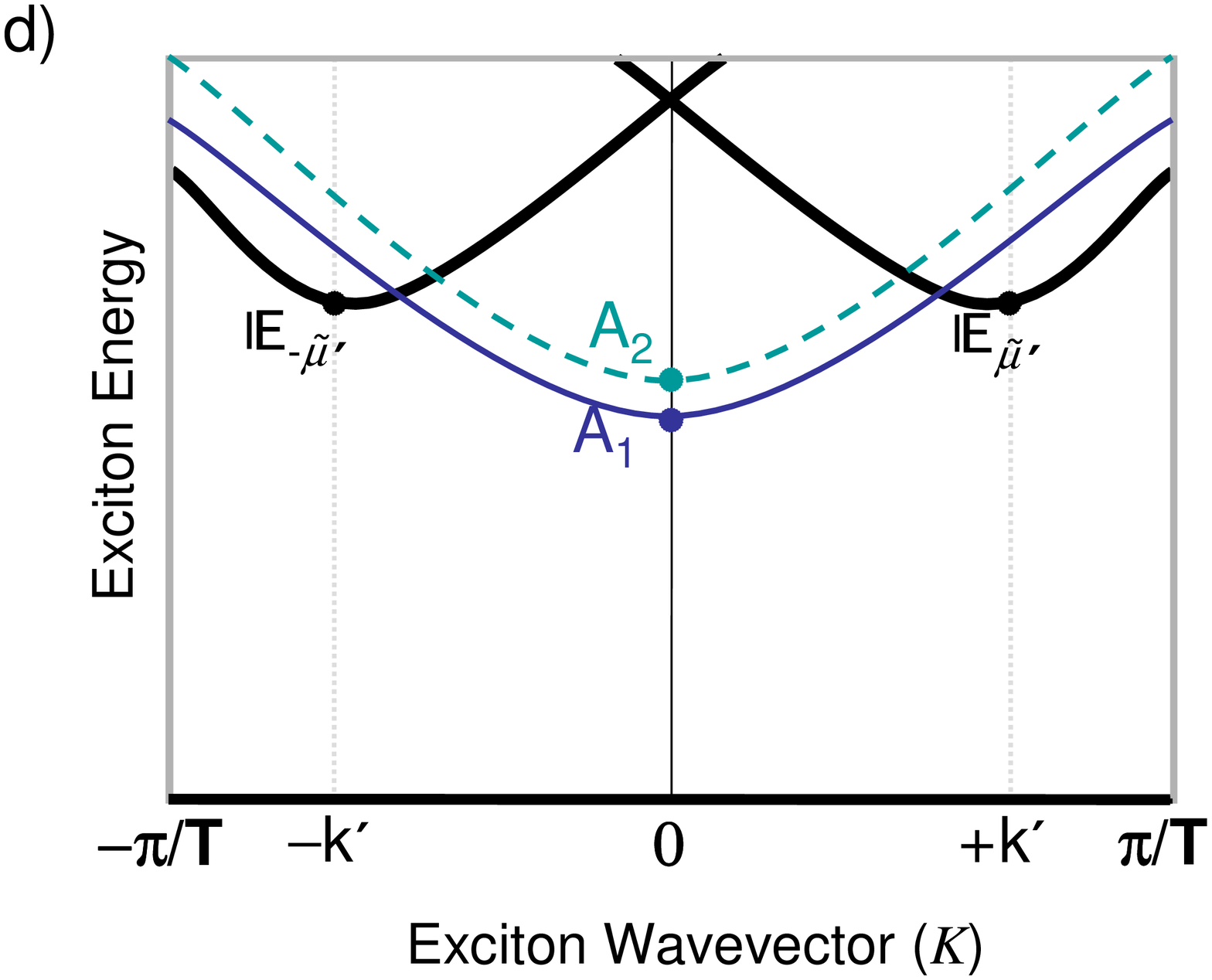}
\includegraphics[width=5.2cm]{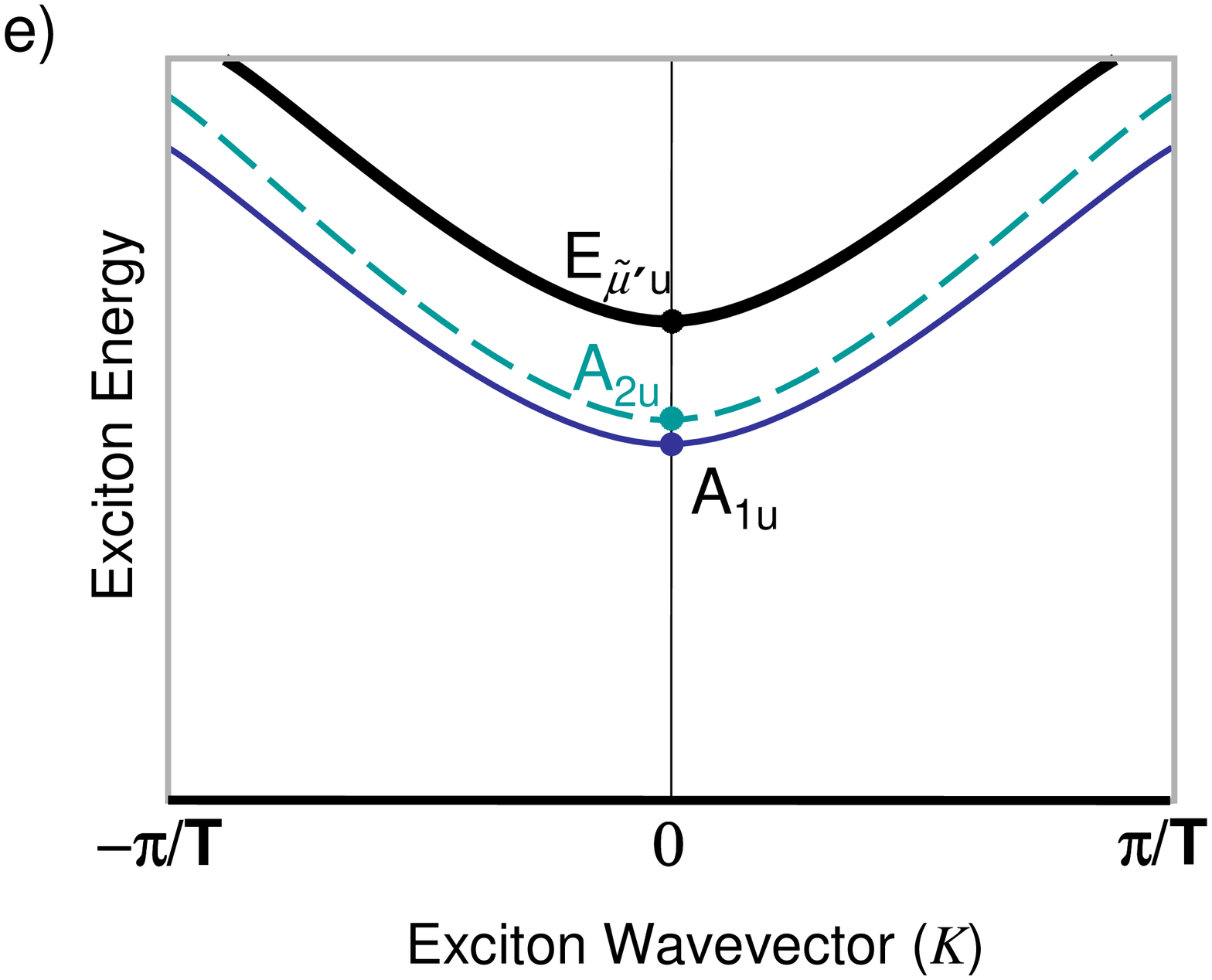}
\includegraphics[width=5.2cm]{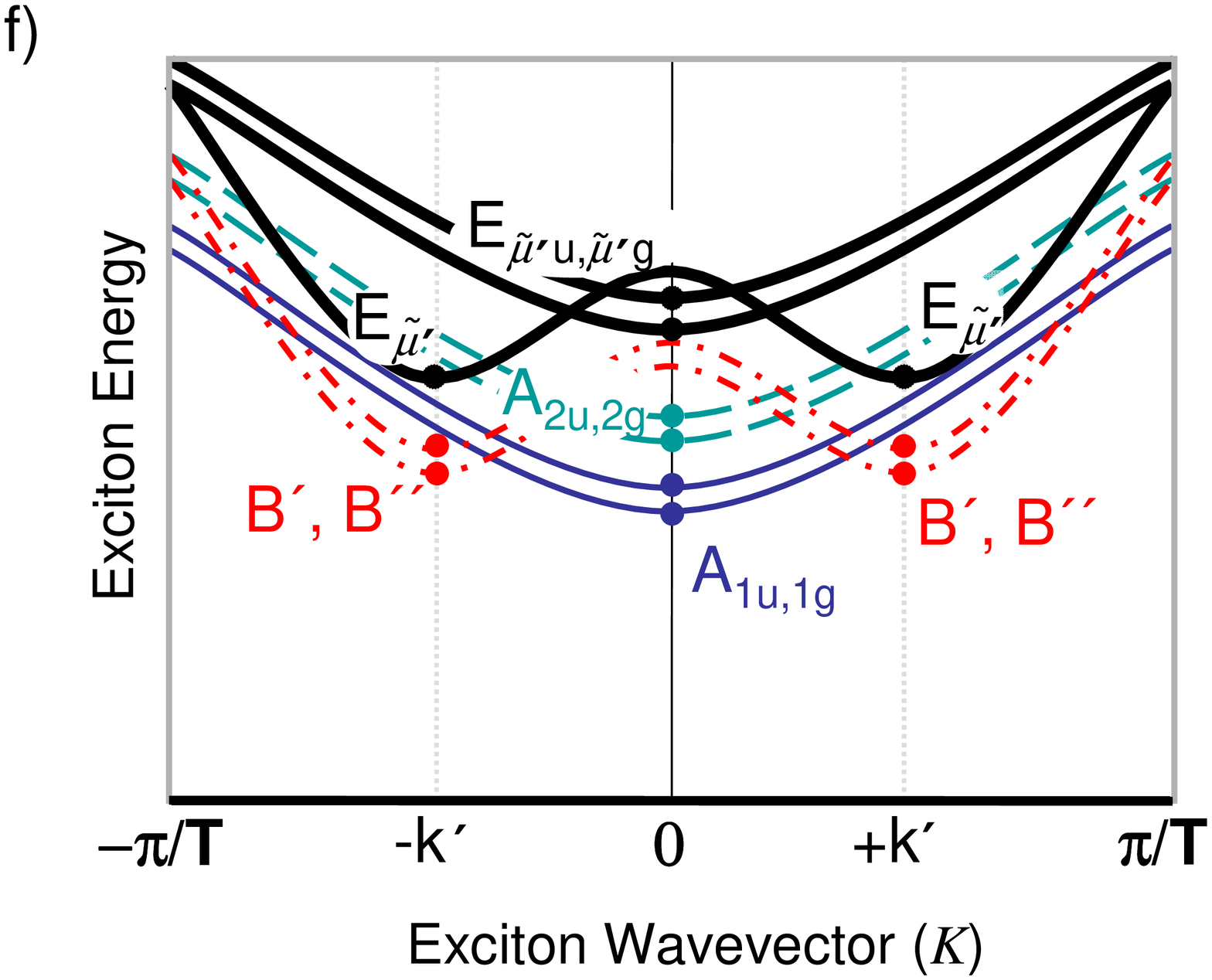}
\caption{{\it Color online} -- Diagrams for the electronic bands and
symmetries for (a) chiral $(n,m)$ (b) zigzag $(n,0)$ and (c)
armchair $(n,n)$ nanotubes and for their respective excitonic bands
(d), (e) and (f). The electron, hole and exciton states at the band
edges are indicated by a solid circle and labeled according to their
irreducible representation. Different line types and colors in this
figure are related to bands with different symmetries. Thick (black)
solid lines correspond to the $E_{\tilde{\mu}'}$ representation, the
blue (thin) solid lines correspond to $A_1$ excitons while the cyan
(thin) dashed lines correspond to the $A_2$ excitonic states. In the
case of achiral nanotubes, we also have inversion and mirror plane
symmetries. For a better visualization, the bands with different
parities under the inversion and mirror planes were grouped together
and appear with the same line color and pattern. In the case of
armchair nanotubes, the bands which transform as the $B'$ and $B''$
representations are shown using a red dot-dash pattern. The
electronic and excitonic band structures shown here are only
pictorial. Group theory does not order the values for the
eigenenergies and energy dispersions.} \label{eband}\end{figure*}

Figures~\ref{eband}(a), (b) and (c) show a schematic diagram of the
electronic valence and conduction single-particle bands with a given
index $|\tilde{\mu}|$, for general chiral, zigzag and armchair
SWNTs, respectively. The electron and hole states at the band-edge
are labelled according to their irreducible
representations.\cite{alon03,barrosreview} The exciton wavefunction
for the one-dimensional (1D) SWNTs can be written as a linear
combination of products of conduction (electron) and valence (hole)
eigenstates as:
\begin{equation}
\psi(\vec{r}_e,\vec{r}_h) =
\sum_{v,c}A_{vc}\phi_{c}(\vec{r}_e)\phi_{v}^*(\vec{r}_h),\label{exc:psi}
\end{equation}
where $v$ and $c$ stand for valence- and conduction- band states,
respectively. For an {\it ab initio} determination of the
coefficients $A_{vc}$, it is necessary to solve a Bethe-Salpeter
equation \cite{chang04,rohlfing00,spataru041}, which incorporates
many-body effects and describes the coupling between electrons and
holes. The many-body Hamiltonian is invariant under the symmetry
operations of the nanotube and therefore each excitonic eigenstate
will belong to an irreducible representation of the space group of
the nanotube. In general the electron-hole interaction will mix
states with all wavevectors and all bands, but for moderately
small-diameter nanotubes ($d_t<1.5~nm$), the energy separation
between singularities in the single-particle 1D JDOS (joint density
of states) is fairly large and it is reasonable to consider, as a
first approximation, that only the electronic bands contributing to
a given 1D singularity will mix to form the excitonic states
\cite{spataru041}. This is the ideal situation to employ the usual
effective-mass and envelope-function approximations (EMA):
\cite{knox}
\begin{equation}
\psi^{EMA}(\vec{r}_e,\vec{r}_h) = \sum_{v,c}{}
^{^{\prime}}B_{vc}\phi_{c}(\vec{r}_e)\phi_{v}^*(\vec{r}_h)F_{\nu}(z_e-z_h).\label{ema}
\end{equation}
The prime in the summation indicates that only those states
associated with the 1D JDOS singularity are included and the
coefficients $B_{vc}$ are dictated by symmetry. It is important to
emphasize that the approximate wavefunctions $\psi^{EMA}$ have the
same symmetries as the full wavefunctions $\psi$. The use of such
"hydrogenic" envelope-functions serves merely as a physically
grounded guess for the ordering in which the different exciton
states appear. The envelope function $F_\nu(z_e-z_h)$ provides an
{\it ad-hoc} localization of the exciton in the relative coordinate
$z_e-z_h$ along the axis and $\nu$ labels the levels in the 1D
hydrogen series \cite{loudon59}. The envelope functions will be
either even ($\nu=0,2,4...$) or odd ($\nu=1,3,5...$) upon
$z\rightarrow -z$ operations. The irreducible representation of the
excitonic state $\mathcal{D}(\psi^{EMA})$ is given by the direct
product:
\begin{equation}
\mathcal{D}(\psi^{EMA}) = \mathcal{D}(\phi_{c})\otimes
\mathcal{D}(\phi_{v})\otimes\mathcal{D}(F_{\nu}),\label{product}
\end{equation}
where $\mathcal{D}(\phi _{c})$, $\mathcal{D}(\phi_{v})$ and
$\mathcal{D}(F_{\nu})$ are the irreducible representations of the
conduction state, valence state and envelope function,
respectively.\cite{knox} We now apply Eq. (\ref{product}) to study
the symmetry of excitons in chiral and achiral (zigzag and armchair)
carbon nanotubes. Let us first consider the first optical transition
(E$_{11}$) in the most general case, the chiral tubes.

{\it Chiral -} As shown in Fig.~\ref{eband}(a), there are two
inequivalent electronic bands in chiral tubes, one with the band
edge at $k=k_0$ and the other one at $k=-k_0$. In order to evaluate
the symmetry of the excitonic states, it is necessary to consider
that the Coulomb interaction will mix the two inequivalent states in
the conduction band (electrons) with the two inequivalent states in
the valence band (holes). These electron and hole states at the vHSs
transform as the 1D representations \cite{notation}
$\mathbb{E}_{\tilde{\mu}}(k_0)$ and
$\mathbb{E}_{-\tilde{\mu}}(-k_0)$ of the $C_N$ point group
\cite{barrosreview}, where we have considered that conduction and
valence band extrema occur at the same $k=k_0$. Taking this into
consideration, the symmetries of the exciton states with the $\nu=0$
envelope function, which transform as the $A_1(0)$ representation,
can be obtained using the direct product in Eq.~(\ref{product}):

\vspace{-0.3cm}
\begin{footnotesize}
\begin{eqnarray} \left(\mathbb{E}_{\tilde{\mu}}(k_0) +
\mathbb{E}_{{-}\tilde{\mu}}({-}k_0)\right)
\otimes\left(\mathbb{E}_{{-}\tilde{\mu}}({-}k_0) +
\mathbb{E}_{\tilde{\mu}}(k_0)\right)\otimes A_1(0) =\nonumber\\
A_1(0) + A_2(0) + \mathbb{E}_{\tilde{\mu}'}(k')+
\mathbb{E}_{{-}\tilde{\mu}'}({-}k'), \label{eq:total}
\end{eqnarray}
\end{footnotesize}
where $k'$ and $\tilde{\mu}'$ are the exciton linear momenta and
quasi-angular momenta, respectively. Note that we considered the
quantum numbers for hole states to be opposite in sign from those of
electron states. Therefore, group theory shows that the lowest
energy set of excitons is composed of four exciton bands, shown
schematically in Fig.~\ref{eband}(d). Basically, the mixing of two
electrons and two holes generates four exciton states. The mixing of
electron and hole states with opposite quantum number $k$ ($k_e=\pm
k_0$, $k_h=\mp k_0$) will give rise to excitonic states which
transform as the $A_1$ and $A_2$ representations of the $D_N$ point
group. These representations correspond, respectively, to states
even and odd under the $C_2$ rotation. These excitons will have a
band minimum at the $\Gamma$ point. The excitonic states formed from
electrons and holes with $k_e=k_h=\pm k_0$ will transform as the
$\mathbb{E}_{\tilde{\mu}'}(k')$ and
$\mathbb{E}_{-\tilde{\mu}'}(-k')$ 1D irreducible representations of
the $C_N$ point group, with an angular quantum number
$\tilde{\mu}'=2\tilde{\mu}$. These exciton states will have a band
edge at $k'=2k_0$ if $2k_0$ is within the 1st Brillouin zone (1BZ).
If $2k_0$ crosses the boundary of the 1st Brillouin zone or
$2\tilde{\mu}$ is larger than $N/2$, the values of $k'$ and
$\tilde{\mu}'$ have to be translated back into the 1st Brillouin
zone.\cite{barrosreview,damn03} It should be mentioned that the
values of $\tilde{\mu}$ and $k_0$ will be different for each
nanotube and also for each E$_{ii}$ transition.

Let us now consider higher-energy exciton states $\nu>0$ for the
same vHSs in JDOS (referred to as $E_{11}$) in chiral tubes. For
$\nu$ even, the resulting decomposition is the same as for $\nu=0$,
since the envelope function also has $A_1$ symmetry. For odd values
of $\nu$, the envelope function will transform as $A_2$, but that
will also leave the decomposition in Eq.(\ref{eq:total}) unchanged.
The result is still the same if one now considers higher-energy
exciton states derived from higher singularities in the JDOS
(E$_{22}$ or E$_{33}$ transitions). Therefore, {\it
Eq.(\ref{eq:total}) describes the symmetries of all exciton states
in chiral nanotubes associated with E$_{ii}$ transitions.}

To obtain the selection rules for the optical absorption of the
excitonic states, it is necessary to consider that the ground state
of the nanotube transforms as a totally symmetric representation
($A_1$) and that only $K=0$ excitons can be created due to linear
momentum conservation. For light polarized parallel to the nanotube
axis, the interaction between the electric field and the electric
dipole moment in the nanotube transforms as the $A_2$ representation
for chiral nanotubes.\cite{barrosreview} Therefore, from the 4
excitons obtained for each envelope function $\nu$, only the $A_2$
symmetry excitons are optically active for parallel polarized light,
the remaining three being dark states. It is clear that the
experimental Kataura plot \cite{bachilo02,z1066} can be interpreted
as the plot of the energy of the bright exciton state with $\nu=0$
as a function of tube diameter. For 2-photon excitation experiments,
the excitons with $A_1$ symmetry are accessed ($A_2 \otimes
A_2=A_1$), and thus, {\it there will also be one bright exciton for
each $\nu$ envelope function}. This result indicates that group
theory does not predict the selection rules used in
Ref.~\onlinecite{wang05}. Thus, the explanation of the results
obtained in 2-photon excitation experiments does not rely on
symmetry selection rules and should be related to oscillator
strength arguments.\cite{maultzsch05} For instance, the bright
exciton associated to odd $\nu$ states in chiral tubes can be
understood as a product between an even Bloch function and an odd
envelope function.\cite{maultzsch05} Therefore, although being
formally bright, we expect a very low oscillation strength for these
excitons, since an odd envelope function should give a very low
probability of finding an electron and a hole at the same position
available for recombination.

\textit{Zigzag -} For zigzag nanotubes, the vHSs for the electronic
bands associated with all $E_{ii}$ transitions occur at $k_0=0$, and
thus, the symmetry of the electron (hole) states will form the
direct product for $\nu$ even:

\vspace{-0.3cm}
\begin{footnotesize}
\begin{equation}
E_{\tilde{\mu}g}(0)\otimes E_{\tilde{\mu}u}(0)\otimes A_{1g}(0)=A_{1u}(0) +
A_{2u}(0) + E_{\tilde{\mu}' u}(0),
\label{eq:zigzag1}
\end{equation}
\end{footnotesize}
and for $\nu$ odd:
\begin{footnotesize}
\begin{equation}
E_{\tilde{\mu}g}(0)\otimes E_{\tilde{\mu}u}(0)\otimes
A_{2u}(0)=A_{2g}(0) + A_{1g}(0) + E_{\tilde{\mu}' g}(0).
\label{eq:zigzag2}
\end{equation}
\end{footnotesize}
The corresponding band structure for $\nu=0$ (lowest exciton states)
is shown in Fig. \ref{eband}(e). It is interesting to note that, in
this case, all four excitonic states will have the band edge at the
$\Gamma$ point ($K=0$). The value of $\tilde{\mu}'$ can be obtained
in the same way as in the case of chiral nanotubes.

For achiral nanotubes, the electromagnetic interaction with the
nanotube transforms as the $A_{2u}$ representation, and thus, one
can see from Eq.(\ref{eq:zigzag1}) and Eq.(\ref{eq:zigzag2}) that
for zigzag nanotubes only states with $\nu$ even (envelope functions
even under $z\rightarrow -z$) will have a bright exciton. Therefore,
group theory predicts that zigzag tubes have a smaller number of
allowed optical transitions than chiral tubes, which is consistent
with their higher symmetry. For 2-photon excitation and emission in
achiral tubes, we have $A_{2u} \otimes A_{2u} = A_{1g}$, and
therefore only the $A_{1g}$ excitons will be optically active. For
zigzag tubes (see Eq.(\ref{eq:zigzag2})), only the states with odd
envelope functions will be accessible by 2-photon transitions, in
agreement with Ref.~\onlinecite{wang05} in this special case.

\textit{Armchair -} The optical transitions in armchair tubes are
also excitonic, despite the metallic character of these tubes,
because of symmetry gap effects \cite{spataru041}. As shown in Fig.
\ref{eband}(c), the $E_{ii}$-derived excitons will be formed by two
$E_{\tilde{\mu}}$ bands at $k=\pm k_0$, where $k_0\approx2\pi/3a$
for the lowest-energy excitons. Therefore, these excitonic states
will be given by the direct product:

\vspace{-0.3cm}
\begin{footnotesize}
\begin{eqnarray}
\left(E_{\tilde{\mu}}(k_0) + E_{\tilde{\mu}}(-k_0)\right)
\otimes\left(E_{\tilde{\mu}}(k_0) + E_{\tilde{\mu}}(-k_0)\right)
\otimes A_{1g,2u} = \nonumber \\ A_{1u}(0) + A_{2u}(0) + A_{1g}(0) +
A_{2u}(0) + \nonumber \\(B'(k')+B'(-k')) + (B''(k')+B''(-k'))
+\nonumber\\ E_{\tilde{\mu}'g}(0) + E_{\tilde{\mu}'u}(0) +
(E_{n-\tilde{\mu}'}(k')+E_{n-\tilde{\mu}'}(-k')). \label{eq:totarm}
\end{eqnarray}
\end{footnotesize}
The same decomposition is found for $A_{1g}$ and $A_{2u}$ envelope
functions. Therefore, each JDOS vHS for armchair SWNTs gives rise to
16 exciton states, as shown in Fig. \ref{eband}(f) for $\nu=0$. If
$k_0=2\pi/3a$, then $k'=k_0$ (the exciton momentum has to be
translated back to the 1st Brillouin Zone). The excitons at $K=0$
transform as the representations of the $D_{2nh}$ group, while the
excitons at $K=\pm k'$ transform as the irreducible representations
of the $C_{2nv}$ point group.

As in the case of zigzag nanotubes, only the $A_{2u}$ ($A_{1g}$)
symmetry exciton will be optically active for 1-photon excitation
(2-photon). Therefore, from the 16 exciton states obtained for each
envelope function $\nu$ there will be one bright exciton. Note that,
in the case of armchair nanotubes, there will also be bright
excitons with odd $\nu$ envelope functions. However, we note that
because of the weak electron-hole interaction due to metallic
screening, the existence of higher $\nu$ states is unlikely in
armchair tubes.

To summarize, we obtained the symmetry of excitonic states in
chiral, zigzag and armchair SWNTs within the approach of the group
of the wavevector $k$. Each set of electronic transitions E$_{ii}$
gives rise to a series of exciton states, each associated with an
envelope function. We show the absence of selection rules for even
and odd envelope functions for most of the carbon nanotubes (i.e.
chiral and armchair). {\it This result shows that group theory does
not predict the one and two-photon selection rules used in the
interpretation of recent experiments.} \cite{wang05} When symmetry
selection rules do not come into play, the existence or apparent
absence of optical transitions should be interpreted in terms of
their high or low oscillator strength.\cite{maultzsch05} It is
important also to stress that zigzag nanotubes are a very special
class of tubes, with very specific symmetry aspects. Generalizing
results from zigzag carbon nanotubes to other symmetry tubes is not
always appropriate.

\begin{acknowledgments} The MIT authors acknowledge
support under NSF Grant DMR04-05538. S.L., S. I-B. C. D. S.
acknowledge financial support from NSF Grant DMR04-39768, U.S.
Department of Energy, Contract No. DE- AC02-05CH11231 and CMSN. E.
B. B., A. G. S. F and A. J. and R. B. C. also acknowledge financial
support from CAPES (PDEE), FUNCAP, CNPq and Faperj Brazilian
agencies, respectivelly. The Brazilian authors acknowledge support
from Instituto de Nanotecnologia and Rede Nacional de Pesquisa em
Nanotubos de Carbono.
\end{acknowledgments}


\begin{thebibliography}{19}
\bibitem[{not()}]{notation}
\bibinfo{note}{The notation $\mathbb{E}$ is used to differentiate this 1D
  representation of the $C_N$ point group from the 2D representation $E$ of the
  higher symmetry point groups $D_N$, $C_{2nv}$ and $D_{2nh}$.}
\end{thebibliography}
\end{document}